\documentclass[3p,times]{elsarticle}

\makeatletter
\def\ps@pprintTitle{%
	\let\@oddhead\@empty
	\let\@evenhead\@empty
	\def\@oddfoot{\centerline{\thepage}}%
	\let\@evenfoot\@oddfoot}
\makeatother

\usepackage{lineno,hyperref}
\usepackage{graphics} 
\usepackage{epsfig} 
\usepackage{amsmath} 
\usepackage{mathrsfs}
\usepackage{amssymb}  
\usepackage{circuitikz}
\usepackage[inline]{enumitem}
\usepackage{etex}
\usepackage{algpseudocode}
\usepackage{wrapfig}
\usepackage{booktabs}
\usepackage{multirow}
\usepackage{scalefnt}
\usepackage{multicol}
\usepackage{scalerel}

\newcommand{\ddt}{\frac{d}{dt}}

\newcommand{\R}{\ensuremath{\mathbb{R}} }
\newcommand{\Nat}{\ensuremath{\mathbb{N}} }
\newcommand{\C}{\ensuremath{\mathbb{C}} }

\newcommand{\Cinf}{\ensuremath{\mathfrak{C}^{\infty}}}
\newcommand{\Cimp}{\ensuremath{\mathfrak{C}_{\rm imp}}}
\newcommand{\n}{{\tt{n}}}
\newcommand{\f}{{\tt{f}}}
\newcommand{\ns}{{\tt{n_s}}}
\newcommand{\nf}{{\tt{n_f}}}
\renewcommand{\r}{{\tt{r}}}
\newcommand{\m}{\tt{m}}
\newcommand{\p}{\tt{p}}
\newcommand{\dd}{\tt{d}}
\newcommand{\N}{\tt{N}}

\newcommand{\Rn}{\R^{\n}}
\newcommand{\Rp}{\R^{\p}}

\newcommand{\Rm}{\R^{\m}}
\newcommand{\Rnn}{\R^{\n \times \n}}

\newcommand{\Rnp}{\R^{\n \times \p}}
\newcommand{\Rmn}{\R^{\m \times \n}}
\newcommand{\Rnm}{\R^{\n \times \m}}

\newcommand{\V}{\mathbb{V}}

\newcommand{\Rea}{\mathcal{R}_s}

\newcommand{\Obs}{\mathcal{O}_w}
\newcommand{\Obsg}{\mathcal{O}_{wg}}

\newcommand{\tA}{\widehat{A}}
\newcommand{\tB}{\widehat{B}}
\newcommand{\tC}{\widehat{C}}
\newcommand{\tD}{\widehat{D}}

\newcommand{\Ns}{\ensuremath{{\tt N}_{\tt s}}}
\newcommand{\M}[1]{\ensuremath{\mathcal{M}_{#1}}}

\newcommand{\del}[1]{\ensuremath{\delta^{(#1)}}}
\newcommand{\Uimp}{\ensuremath{\mathcal{U}_{\tt imp}}}
\newcommand{\hN}{\ensuremath{\widehat{N}}}

\newcommand{\bA}{A}
\newcommand{\bB}{B}
\newcommand{\bC}{C}
\newcommand{\bD}{D}

\newcommand{\g}{\tt g}
\newcommand{\w}{\tt w}
\renewcommand{\S}{\mathcal{S}}

\newtheorem{theo}{Theorem}[section]
\newtheorem{definition}[theo]{Definition}

\newtheorem{lemma}[theo]{Lemma}

\newtheorem{proposition}[theo]{Proposition}
\newtheorem{remark1}[theo]{Remark}
\newenvironment{remark}{\begin{remark1} \rm } {\rm \hfill $\Box$\end{remark1} }
\newenvironment{proof1}{\noindent {\em Proof}\ }{\hfill $\Box$ \vspace{1mm}\newline}

\begin{document}

\begin{frontmatter}

\title{A matrix theoretic characterization of the strongly reachable subspace}

\author{Imrul Qais}
\address{Indian Institute of Technology Bombay, Mumbai, India}
\ead{imrul@ee.iitb.ac.in}

\author{Chayan Bhawal}
\address{Indian Institute of Technology Guwahati, Guwahati, India}
\ead{bhawal@iitg.ac.in}

\author{Debasattam Pal}
\address{Indian Institute of Technology Bombay, Mumbai, India}
\ead{debasattam@ee.iitb.ac.in}

\begin{abstract}
In this paper, we provide novel characterizations of the weakly unobservable and the strongly reachable subspaces corresponding to a given state-space system. These characterizations provide closed-form representations for the said subspaces. In this process we establish that the strongly reachable subspace is intimately related to the space of admissible impulsive inputs. We also show how to calculate the dimensions of these subspaces from the transfer matrix of the system.
\end{abstract}

%

\end{frontmatter}

\section{Introduction}

One of the most interesting ideas that had been put forward in linear system theory 
has been the notion of invariant subspaces, in particular the idea of $A \,{\tt mod}B$ 
invariant subspaces and controllability subspaces \cite{Wil:80},\cite{Won:85}. These notions 
have not only enriched the theory on linear systems by making the `fine structure' of 
multivariable linear systems apparent, but, more importantly, they have also been 
instrumental in solving a wide variety of control theoretic questions like 
disturbance decoupling, output stabilization, tracking and regulation, decoupling, etc 
\cite{Won:85}. It was in \cite{Wil:80} that the authors used the notion of 
{\em \underline{almost} $A \,{\tt mod}B$ invariant subspaces} and {\em \underline{almost} 
	controllability subspaces} in order to answer the important question in disturbance 
decoupling problems; if there exist feedback matrices such that the effect of noise on 
the output is small. This question was relevant, since the idea of $A \,{\tt mod}B$ invariant 
subspaces and controllability subspaces failed to provide any solution for disturbance decoupling when  
certain conditions are not met. The introduction of the almost invariant subspaces in 
\cite{Wil:80} led to the idea of weakly unobservable subspaces and strongly reachable subspaces in 
control theory. It is in \cite{HauSil:83} that these ideas were used to solve optimal control 
problems. These ideas resurfaced in the literature again in the study of linear complementarity systems 
(LCS) \cite{HeeSchWei:00}. In LCS, the weakly unobservable system corresponds to the consistent 
subspace and the the strongly reachable subspace of the system are known as the fast subspace. In 
\cite{HeeSchWei:00}, it was shown that LCS viewed as a collection of linear systems switching between 
operating points require the notion of fast subspaces for the characterization of jumps in states. 
The consistent and fast subspace of a linear system become essential in the observer design for LCS, 
as well \cite{HeeKanSchJoh:11}. These subspaces have therefore played important role in different areas 
of control theory. Hence, it is natural to ask questions like - how do we compute a basis corresponding to these 
subspaces? or how do we find the dimensions of these subspaces? Iterative algorithms 
in \cite{HauSil:83} answer these questions to some extent. However, these algorithms do not reveal the fine 
structure of the subspaces. For example, a basis for strongly reachable subspace computed using the iterative 
algorithm in \cite{HauSil:83} do not reveal any information as to how this basis is linked to the 
system matrices. Further, the iterative algorithms reveal the dimension of the weakly unobservable 
and strongly reachable subspaces, once the algorithms terminate. Can these dimensions be known from the system 
matrices directly? Weakly unobservable subspace of a system being associated with the deflating subspaces of 
the corresponding matrix pencil is another aspect which, although known to many researchers in this area, have 
not been explored much in the literature. Hence, in this paper, we provide an extensive geometric characterization 
of the subspaces. This leads us to new algorithms to compute these subspaces. These algorithms use fundamental linear 
algebraic notions of nullspace and eigenspace of suitable matrices formed using the system matrices. 
Another important problem in system theory is the output-nulling problem. The problem finds widespread application 
in optimal control problems related to standard state-space systems, implicit systems, differential algebraic systems, 
etc \cite{Won:85}, \cite{Ter:98}, \cite{KazNtoPer:18}. Over the years there has been substantial research on this 
problem as well \cite{And:75}, \cite{Fuh:05}, \cite{PadNto:20}. Recently, in \cite{PadFerNto:21} the authors characterized
the output nulling subspaces obtained from the kernels of Rosenbrock pencils in terms of the reachability subspaces. 
Output nulling subspaces are intrinsically linked to the notion of weakly unobservable subspaces. The characterization 
of weakly unobservable subspace have already been reported in \cite{QaiBhaPal:20}. The same has been presented in 
this paper for the sake of completeness. In this paper, we primarily focus on the characterization of the strongly 
reachable subspace. To the best of our knowledge such a characterization has not been established in the literature yet. 

\section{Notation and Preliminaries}\label{LAAsec:prelims}

\subsection{Notation}
The symbols $\R$, $\C$, and $\Nat$ are used for the sets of real numbers, complex numbers, and natural numbers, respectively. We use the symbol $\R_+$ and $\C_-$ for the sets of positive real numbers and complex numbers with negative real parts, respectively. $\R[s]$ and $\R(s)$ denote the ring of polynomials with real coefficients and the field of rational functions, respectively. The symbols $\Rnp,\R[s]^{\n\times\p},$ and $\R(s)^{\n\times\p}$ denote the set of ${\n} \times {\p}$ matrices with elements from $\R,\R[s],$ and $\R(s)$, respectively. We use $\bullet$ when a dimension need not be specified: for example, $\R^{{\w} \times \bullet}$ denotes the set of real constant matrices having ${\w}$ rows and an unspecified number of columns. We use the symbol $I_{\n}$ for an ${\n} \times {\n}$ identity matrix and the symbol $0_{{\n},{\m}}$ for an ${\n} \times {\m}$ matrix with all entries zero. The symbol $\{0\}$ is used to denote the zero subspace. Symbol ${\tt col}(B_1, B_2,\! \ldots,\! B_{\n})$ represents a matrix of the form $\left[\begin{smallmatrix}B_1^T & B_2^T & \cdots & B_{\n}^T\end{smallmatrix}\right]^T$. The symbol ${\tt img\,}A$ and ${\tt ker\,}A$ denote the image and nullspace of a matrix $A$, respectively. The symbol ${\tt det}(A)$ represents the determinant of a square matrix $A$. Symbol ${\tt num}(p(s))$ is used to denote the numerator of a rational function $p(s)$. A combination of these two symbols ${\tt num}{\tt det}A(s)$ denotes the numerator of the determinant of a rational function matrix $A(s)$. We use the symbol ${\tt dim\,}(\S)$ to denote the dimension of a space $\S$. The space of all infinitely differentiable functions from $\R$ to $\Rn$ is represented by the symbol $\Cinf(\R,\Rn)$. We use the symbol $\Cinf(\R,\Rn)|_{\R_+}$ to represent the set of all functions from $\R_{+}$ to $\Rn$ that are restrictions of $\Cinf(\R,\Rn)$ functions to $\R_{+}$. The symbol $\delta$ represents the Dirac delta impulse distribution and $\delta^{(i)}$ represents the $i$-th distributional derivative of $\delta$ with respect to $t$.

\subsection{The weakly unobservable and the strongly reachable subspaces}\label{LAAsec:weak_strong}
Consider the system $\Sigma$ with an input-state-output (i/s/o) representation:
\begin{equation}{\label{eq:sys_sig}}
\ddt x=Ax+Bu \text{ and } y=Cx+Du, 
\end{equation} 
where $A \in \Rnn, B \in \Rnm$, $C \in \R^{\p \times \n}$ and $D\in\R^{\p\times\m}$. Associated with such a system are two important 
subspaces called the {\em weakly unobservable} subspace (or, the {\em slow space}) and the {\em strongly reachable} subspace 
(or, the {\em fast space}). Before we delve into the definitions of these subspaces, we need to define the space of impulsive-smooth 
distributions (see \cite{HauSil:83}, \cite{WilKitSil:86}).
\begin{definition}\label{LQRdef:Cimp}
	The set of impulsive-smooth distributions $\Cimp^{\m}$ is defined as:
	\begin{align*}		
	\Cimp^{\m} := \Big\{f= f_{\tt reg} + f_{\tt imp}\,|\, f_{\tt reg} 
	\in \Cinf(\R,\Rm)|_{\R_+} \mbox{ and } f_{\tt imp} = \sum_{i=0}^k a_i\delta^{(i)}, \mbox{ with } a_i
	\in \Rm, k \in 	\Nat \Big\}.
	\end{align*}
\end{definition}
In what follows, we use symbols $x(t;x_0,u)$ and $y(t;x_0,u)$, respectively, to denote the state-trajectory $x$ and 
the output-trajectory $y$ of the system $\Sigma$, that result from initial condition $x_0$ and input $u(t)$. 
The symbol $x(0^+;x_0,u)$ denotes 
the value of the state-trajectory that can be reached from $x_0$ instantaneously on application of the input $u(t)$ at $t = 0$.
\begin{definition}\label{LQRdef:weakly}
	Consider the system $\Sigma$ defined in equation \eqref{eq:sys_sig}. A state $x_0 \!\in\! \Rn$ is called {\em weakly unobservable} 
	if there exists an input $u \in \Cinf(\R,\Rm)|_{\R_+}$ such that $y(t;x_0,u) \equiv 0$ for all $t \geqslant 0$. The collection of 
	all such weakly unobservable states is a subspace. This subspace is called the {\em weakly unobservable subspace} (or, the 
	{\em slow space}) of the state-space.
\end{definition}
The following proposition from \cite{HauSil:83} gives a geometric interpretation of the weakly unobservable subspace.
\begin{proposition}\label{LQRprop:property_weakly}
	The weakly unobservable subspace $\Obs$ is the largest subspace $\mathcal{V}$ of the state-space for which there exists a 
	feedback $F\in\Rmn$ such that
	\begin{eqnarray}\label{eqn:property_weakly}
	(A+BF)\mathcal{V}\subseteq\mathcal{V} \text{ and } (C+DF)\mathcal{V}=0.
	\end{eqnarray}
	In other words, if $\mathcal{V}$ is any subspace that satisfies equation\eqref{eqn:property_weakly}, then $\mathcal{V}\subseteq\Obs$.
\end{proposition}
An important subspace of the weakly unobservable subspace is widely used in various problems like linear quadratic regulator (LQR) 
problem and KYP lemma. We call this subspace the {\em good weakly unobservable} subspace. Following is the formal definition of this subspace.
\begin{definition}\label{def:good_slow_space}
	The good weakly unobservable subspace $\Obsg$ is the largest subspace $\mathcal{V}$ of the state-space for which there exists a feedback 
	$F\in\Rmn$ such that
	\begin{eqnarray}\label{eqn:good_slow_space_defn}
	(A+BF)\mathcal{V}\subseteq\mathcal{V},~ (C+DF)\mathcal{V}=0, \mbox{ and } \sigma((A+BF)|_{\mathcal{V}})\subseteq \C_{-}.
	\end{eqnarray}
	In other words, if $\mathcal{V}$ is any subspace that satisfies equation \eqref{eqn:good_slow_space_defn}, then $\mathcal{V}\subseteq\Obsg$.
\end{definition}

Properties of the weakly unobservable and the good weakly unobservable subspaces have been discussed in detail in \cite{HauSil:83} and 
\cite{QaiBhaPal:20}. In this paper we deal with the strongly reachable subspace. But, for the sake of completeness, we present some of the 
main results from \cite{QaiBhaPal:20} in Section \ref{LAAsec:char_slow_space}. The definition of the strongly reachable subspace is 
presented next.
\begin{definition}\label{LQRdef:strong}
	Consider the system $\Sigma$ defined in equation \eqref{eq:sys_sig}. A state $x_1\! \in\! \Rn$ is called {\em strongly reachable} 
	(from the origin) if there exists an input $u(t) \in \Cimp^{\m}$ such that $x(0^+;0,u)\!=\!x_1$ and 
	$y(t;0,u)\!\in\! \Cinf(\R,\Rp)|_{\R_+}$. The collection of all such strongly reachable states is a subspace. This subspace is called the 
	{\em strongly reachable subspace} ({\em or} the fast space) of the state-space and is denoted by $\Rea$.
\end{definition}
In \cite{HauSil:83}, the following recursive algorithm is given to compute the subspace $\Rea$:
\begin{align}\label{LQReqn:fast_recursion}
\mathcal{R}_0 := \{0\} \subsetneq \R^{\n}, \mbox{ and } \mathcal{R}_{i+1} := 
\begin{bmatrix} A & B\end{bmatrix}\left\{(\mathscr{W}_i \oplus \mathscr{P}) \cap {\tt ker}
\begin{bmatrix}C & D\end{bmatrix}\right\} \subseteq \Rea, 
\end{align}	
where $\mathscr{W}_i := \left\{\left[\begin{smallmatrix}w\\0\end{smallmatrix}\right] \in \R^{{\n}+{\m}}\,|\, 
w \in \mathcal{R}_i\right\}$ and $\mathscr{P} := 
\left\{\left[\begin{smallmatrix}0\\\alpha\end{smallmatrix}\right] \in \R^{{\n}+{\m}}\,|\, 
\alpha \in \Rm\right\}$. If $\mathcal{R}_{i}=\mathcal{R}_{i+1}$ for some $i\in\Nat$, then $\Rea=\mathcal{R}_{i}$. In Section 
\ref{LAAsec:sub_char_fast_space} we use this recursive algorithm to provide a closed-form expression for $\Rea$. The subspace $\Rea$ is 
closely related with the space of the {\em admissible impulsive inputs}. We define this space next.
\begin{definition}\label{LQRdef:ad_imp_inp}
	An input $u(t):=\sum\limits_{i=0}^{k}u_{i}\delta^{(i)}$, where $u_{i}\in\Rm$ is called an {\em admissible impulsive input} for $\Sigma$ 
	(defined in equation \eqref{eq:sys_sig}) if $y(t;0,u) \in \Cinf(\R,\Rp)|_{\R_+}$. The collection $\mathcal{U_{\tt imp}}$ of all admissible 
	impulsive inputs is a vector space. We call $\mathcal{U_{\tt imp}}$ the space of admissible impulsive inputs. Further, $\delta^{(k)}$ is 
	said to be admissible in the input if $u(t)$ is an admissible impulsive input with $u_{k}\neq 0$.
\end{definition}
From Definition \ref{LQRdef:ad_imp_inp}, the following proposition follows immediately.
\begin{proposition}\label{LQRprp:ad_imp_inp}
	Consider $u(t)$ as defined in Definition \ref{LQRdef:ad_imp_inp}. Then, $u(t)\in\Uimp$ if and only if $G(s)U(s)$ is strictly proper 
	(see Remark \ref{rem:proper_matrices}), where $U(s):=\sum\limits_{i=0}^{k}u_{i}s^{i}$ and $G(s):=C(sI_{\n}-A)^{-1}B+D$ is the transfer 
	matrix of~ $\Sigma$.	
\end{proposition}
\begin{remark}\label{rem:proper_matrices}
	A matrix $A(s)\in\R(s)^{\r\times \tt{c}}$ is called {\em (strictly) proper} if each of the entries of $A(s)$ is (strictly) proper.
\end{remark}

\section{Characterization of the weakly unobservable subspace}\label{LAAsec:char_slow_space}
\noindent In this section, we characterize the weakly unobservable subspace of the system $\Sigma$ as defined in equation \eqref{eq:sys_sig}. 
As mentioned earlier, these result has already been published in \cite{QaiBhaPal:20}. We include these results in this paper for the sake of 
completeness. It should be noted that \cite{QaiBhaPal:20} assumes that the system $\Sigma$ is square, that is, the input-cardinality and the 
output-cardinality of $\Sigma$ are equal. Therefore, $B,C^T\in\Rm$. This characterization is achieved in terms of an eigenspace of the Rosenbrock 
matrix pair $(U_1,U_2)$ defined as follows:
\begin{eqnarray}\label{eqn:ros_sys_mat}
U_1:=\left[\begin{smallmatrix}
I_{\n} & 0\\
0 & 0
\end{smallmatrix}\right]\in\R^{(\n+\m)\times(\n+\m)} \mbox{ and } U_2:=\left[\begin{smallmatrix}
A & B\\
C & D
\end{smallmatrix}\right].
\end{eqnarray}
The following theorem from \cite{QaiBhaPal:20} characterizes the weakly unobservable subspace of $\Sigma$ in terms of an eigenspace of the 
matrix pair $(U_1,U_2)$.
\begin{theo}\label{thm:slow_space}
	Consider the system $\Sigma$ defined in equation \eqref{eq:sys_sig} (with an additional assumption that $\p=\m$) and the corresponding 
	Rosenbrock matrix pair $(U_1,U_2)$ as defined in equation \eqref{eqn:ros_sys_mat}. Assume that ${\tt det}(sU_1-U_2)\neq 0$ and 
	${\tt deg}$${\tt det}(sU_1-U_2)=:\ns$. Let $V_1\in\R^{\n\times \ns}$ and $V_2\in\R^{\m\times\ns}$ be such that ${\tt col}(V_1,V_2)$ is full 
	column-rank and 
	\begin{eqnarray}\label{eqn:slow_space_eigv}
		\underbrace{\begin{bmatrix}
			A & B\\
			C & D
			\end{bmatrix}}_{U_2}\begin{bmatrix}
		V_1\\V_2
		\end{bmatrix}=\underbrace{\begin{bmatrix}
			I_{\n} & 0\\
			0 & 0
			\end{bmatrix}}_{U_1}\begin{bmatrix}
		V_1\\V_2
		\end{bmatrix}J,
		\end{eqnarray}
	where $J\in\R^{\ns\times\ns}$ and ${\tt det}(sI_{\ns}-J)={\tt det}(sU_1-U_2)$. Let $\Obs$ be the slow space of $\Sigma$. Then, the following statements hold:
	\begin{enumerate}
		\item $V_1$ is full column-rank.
		\item $\Obs ={\tt img}V_1.$
		\item ${\tt dim}(\Obs)=\ns$.
	\end{enumerate}
\end{theo}
As discussed earlier, an important subspace of the weakly unobservable subspace is the good weakly unobservable subspace. In \cite{QaiBhaPal:20}, this subspace has been characterized in terms of an stable eigenspace of the Rosenbrock matrix pair $(U_1,U_2)$. We present this as a lemma next.
\begin{lemma}\label{lem:good_slow_space}
	Consider the system $\Sigma$ and the corresponding Rosenbrock matrix pair $(U_1,U_2)$ as defined in equation \eqref{eq:sys_sig} and equation \eqref{eqn:ros_sys_mat}, respectively. Assume that ${\tt det}(sU_1-U_2)\neq 0$ and $\n_{\g}:=|{\tt roots}({\tt det}(sU_1-U_2))\cap\C_{-}|$. Let $V_{1g}\in\R^{\n\times \n_{\g}}$ and $V_{2g}\in\R^{\m\times\n_{\g}}$ be such that ${\tt col}(V_{1g},V_{2g})$ is full column-rank and 
	\begin{eqnarray}\label{eqn:slow_space_eigv}
	\underbrace{\begin{bmatrix}
		A & B\\
		C & D
		\end{bmatrix}}_{U_2}\begin{bmatrix}
	V_{1g}\\V_{2g}
	\end{bmatrix}=\underbrace{\begin{bmatrix}
		I_{\n} & 0\\
		0 & 0
		\end{bmatrix}}_{U_1}\begin{bmatrix}
	V_{1g}\\V_{2g}
	\end{bmatrix}J_{g},
	\end{eqnarray}
	where $J_{g}\in\R^{\n_{\g}\times\n_{\g}}$ and ${\tt roots}({\tt det}(sI_{\n_{\g}}-J_{g}))={\tt roots}({\tt det}(sU_1-U_2))\cap\C_{-}$. Let $\Obsg$ be the slow space of $\Sigma$. Then, the following statements hold:
	\begin{enumerate}
		\item $V_{1\g}$ is full column-rank.
		\item $\Obsg={\tt img}V_{1\g}$.
		\item ${\tt dim}(\Obsg)=\n_{\g}$.
	\end{enumerate}
\end{lemma}

\section{Characterization of the strongly reachable subspace}\label{LAAsec:char_fast_space}
\noindent In this section we characterize the space of admissible impulsive inputs $(\Uimp)$ and the strongly reachable subspace $(\Rea)$ for the system $\Sigma$ defined in equation \eqref{eq:sys_sig}. As mentioned earlier, $\Uimp$ and $\Rea$ are very closely related. In the first part of this section, we explicitly establish this relation between them. In the second part, we show how to obtain the dimension of the strongly reachable subspace from the transfer matrix of a given system.
\subsection{Characterization of the space of admissible impulsive inputs and the strongly reachable subspace}\label{LAAsec:sub_char_fast_space}
Suppose, $u(t)\in\Uimp$, where $u(t):=\sum_{i=0}^{k-1}u_{i}\delta^{(i)}$ and $u_{i}\in\Rm$. Also, assume that corresponding to the initial condition $x(0)=0$, the state trajectory resulting from the input $u(t)$ is $x(t)=h(t)+\sum_{i=0}^{k-2}x_{i}\delta^{(i)}$, where $h(t)\in\Cinf(\R,\Rn)|_{\R_+}$ is the regular part of the state trajectory. Then, following \cite[Section 3]{HauSil:83} we determine the coefficients $x_0,x_1,\dots,x_{k-2}$ of the impulsive part of the state trajectory to be
\begin{align}
x_{k-\ell}=Ax_{k-\ell+1}\!+\!Bu_{k-\ell+1}\!=\!A^{(\ell-2)}Bu_{k-1}\!+\!A^{(\ell-3)}Bu_{k-2}\!+\!\dots+Bu_{k-\ell+1}~ \mbox{ for } \ell\in\{2,3,\dots,k\}.\label{eqn:impulsive_part}
\end{align}
Next, the output of the system is given by
\begin{align*}
y(t)=Cx(t)+Du(t)=Ch(t)+\sum_{i=0}^{k-2}(Cx_{i}+Du_{i})\delta^{(i)}+Du_{k-1}\delta^{(k-1)}.
\end{align*}
Since $u(t)\in\Uimp$, we must have that $y(t)\in\Cinf(\R,\Rp)|_{\R_+}$. Therefore,
\begin{align}\label{eqn:output_impulse_free}
\sum_{i=0}^{k-2}(Cx_{i}+Du_{i})\delta^{(i)}+Du_{k-1}\delta^{(k-1)}=0 \Leftrightarrow Du_{k-1}=0 \mbox{ and } Cx_{i}+Du_{i}=0 \mbox{ for }i\in\{0,1,\dots,k-2\}.
\end{align}
Using equation \eqref{eqn:impulsive_part} in equation \eqref{eqn:output_impulse_free}, we conclude that
\begin{align}\label{eqn:rows_of_markov_matrix}
Du_{k-1}=0 \mbox{ and }Du_{k-\ell}+CBu_{k-\ell+1}\!+\!CABu_{k-\ell+2}\!+\!\dots+CA^{(\ell-2)}Bu_{k-1}=0 \mbox{ for }\ell\in\{2,3,\dots,k\}.
\end{align}
Equation \eqref{eqn:rows_of_markov_matrix} can be written in block-matrix form as
\begin{eqnarray}{\label{eqn:markov_para_matrix}}
\underbrace{\begin{bmatrix}
0 & 0 & \dots & 0 & \bD\\
0 & 0 & \dots & \bD & \bC\bB\\
0 & 0 & \dots & \bC\bB & \bC\bA\bB\\
\vdots & \vdots & \ddots & \vdots & \vdots\\
\bD & \bC\bB & \dots & \bC\bA^{k-3}\bB & \bC\bA^{k-2}\bB
\end{bmatrix}}_{\M{k}}\begin{bmatrix}
u_0\\u_1\\\vdots\\u_{k-1}
\end{bmatrix}=0. ~\mbox{ (Note: $\M{k}=D$, for $k=1$)}.
\end{eqnarray}
We call the matrix $\M{k}$, the {\em Markov parameter matrix}. Clearly, $u(t)\in\Uimp\Rightarrow {\tt col}(u_0,u_1,\dots,u_{k-1})\in\M{k}$. It can be easily shown that the converse is also true. It also turns out that $x_{s}=\left[\begin{smallmatrix}
B & AB & \dots & A^{k-1}B
\end{smallmatrix}\right]{\tt col}(u_0,u_1,\dots,u_{k-1})\in\Rea$. We write these results next as a lemma and a theorem, respectively.
\begin{lemma}{\label{lem:ad_imp_dimention}}
	Consider the system $\Sigma$ and the matrix $\mathcal{M}_{k}$ given by equation \eqref{eq:sys_sig} and equation \eqref{eqn:markov_para_matrix}, respectively. Define ${\tt d}:={\tt dim}({\tt ker}\bD).$ Then: 
	\begin{enumerate}
		\item[{\rm 1.}] The following are equivalent:
	\begin{enumerate}
		\item[{\rm (a)}] Dimension of the space of admissible impulsive inputs, $\mathcal{U}_{\tt imp}$, is $\f$.
		\item[{\rm (b)}] ${\tt dim}({\tt ker}\mathcal{M}_{{\tt f}})={\tt dim}({\tt ker}\mathcal{M}_{{\tt f+1}})=\f.$
		\item[{\rm (c)}] ${\tt dim}({\tt ker}\mathcal{M}_{{\tt f-d+1}})={\tt dim}({\tt ker}\mathcal{M}_{{\tt f-d+2}})=\f.$
	\end{enumerate}
       \item[{\rm 2.}] If ${\tt dim}(\Uimp)=\f$, then $\Uimp=\Delta$, where $\Delta:=\bigg\{\sum_{\tt i=0}^{\tt f-d}u_{\tt i}\del{i},u_{\tt i}\in\R^{\tt m}~\vert~\left[\begin{smallmatrix}
       u_0\\u_1\\\vdots\\u_{\tt f-d}
       \end{smallmatrix}\right]\in{\tt ker}\mathcal{M}_{{\tt f-d+1}}\bigg\}.$
	\end{enumerate}
\end{lemma}
The following theorem characterizes the strongly reachable subspace, $\Rea$, of the system $\Sigma$ described by equation \eqref{eq:sys_sig}. This characterization of the strongly reachable subspace is an alternative representation of the recursive algorithm given in \cite{HauSil:83}. But, the result presented here establishes a direct relation between the strongly reachable subspace and the space of admissible impulsive inputs. Furthermore, the algorithm given in \cite{HauSil:83} is a recursive algorithm, whereas the method presented here gives a closed-form expression for $\Rea$. To the best of our knowledge, the existing results do not give any method to compute the dimension of $\Rea$, whereas in this paper we show how to calculate its dimension from the transfer matrix of a given system.
\begin{theo}{\label{LQRthm:fast_space}}
	Consider the system $\Sigma$ and the matrix $\mathcal{M}_{{\tt f-d+1}}$ as described by equation \eqref{eq:sys_sig} and equation \eqref{eqn:markov_para_matrix}, respectively. If ${\tt dim}(\Uimp)=\f$, then the following statements are true:
	\begin{enumerate}
		\item[{\rm 1.}] $\Rea={\tt img}\left[\begin{matrix}
		\bB & \bA\bB & \dots & \bA^{\f-\tt d}\bB
		\end{matrix}\right]N$, where the columns of $N$ form a basis for ${\tt ker}\mathcal{M}_{{\tt f-d+1}}$ and $\dd:={\tt dim}({\tt ker}$$\bD)$.
		\item[{\rm 2.}] ${\tt dim}(\Rea)=\f$.
	\end{enumerate}
\end{theo}
We defer the proofs of Lemma \ref{lem:ad_imp_dimention} and Theorem \ref{LQRthm:fast_space} until we prove two crucial lemmas. Evidently, the matrix $\M{k}$ defined in equation \eqref{eqn:markov_para_matrix} plays a pivotal role in characterization of $\Uimp$ and $\Rea$. In the following lemma we explore some interesting properties of the matrix $\M{k}$.
\begin{lemma}{\label{lem:markov_mat_prop}}
	Consider the matrix $\mathcal{M}_{k}$ as defined in equation \eqref{eqn:markov_para_matrix}. Then the following statements hold:
	\begin{enumerate}
		\item[{\rm 1.}] If $v\in{\tt ker}\mathcal{M}_{k}\subseteq\R^{k\m}$, then 
		${\tt col}(v,0)\in{\tt ker}\mathcal{M}_{k+1}\subseteq\R^{(k+1)\m}.$
		\item[{\rm 2.}] ${\tt dim}({\tt ker}\mathcal{M}_{k+1})\geqslant {\tt dim}({\tt ker}\mathcal{M}_{k})$ for all $k\in\Nat$.
		\item[{\rm 3.}] If ${\tt r}:={\tt dim}({\tt ker}\mathcal{M}_{{\tt i}})={\tt dim}({\tt ker}\mathcal{M}_{{\tt i+1}})$ for some $\tt{i}\in\Nat$, then ${\tt dim}({\tt ker}\mathcal{M}_{k})={\tt r} \mbox{ for all } k\geqslant {\tt i}.$
	\end{enumerate}
\end{lemma}
\begin{proof1}
	\textbf{1.} It can be easily seen that 
	\begin{eqnarray}\label{eqn:part_markov_matrix}
	\mathcal{M}_{k+1}\!\!=\!\!\left[\begin{smallmatrix}
	\!0 & \bD\!\\\!\mathcal{M}_{k} & m_{k+1}\!
	\end{smallmatrix}\right]\!\!=\!\!\left[\begin{smallmatrix}
	0 & \mathcal{M}_{{k}}\\\bD & \ell_{k+1}
	\end{smallmatrix}\right],
	\end{eqnarray}
	$\mbox{ where } m_{k+1}\!\!=\!\! \left[\begin{smallmatrix}
	\bC\bB\\ \bC\bA\bB\\ \vdots\\ \bC\bA^{k-1}\bB
	\end{smallmatrix}\right], \ell_{k+1}\!\!:=\!\!\left[\begin{smallmatrix}
	\bC\bB & \bC\bA\bB & \dots & \bC\bA^{k-1}\bB
	\end{smallmatrix}\right].$ Since $\mathcal{M}_{k}v=0$, it follows that 
	$
	\mathcal{M}_{k+1}\begin{bmatrix}
	v\\0
	\end{bmatrix}=\begin{bmatrix}
	0 & \bD\\
	\mathcal{M}_{k} & m_{k+1}
	\end{bmatrix}\begin{bmatrix}
	v\\0
	\end{bmatrix}=0.
	$
	This proves Statement $1$.
	
	\noindent \textbf{2.} This statement is a direct consequence of Statement $1$.
	
	\noindent \textbf{3.} We prove this statement by induction.\\
	\textbf{Base case ($k={{\tt i}+2}$):} It is given that ${\tt r}={\tt dim}({\tt ker}\mathcal{M}_{{\tt i}})={\tt dim}({\tt ker}\mathcal{M}_{{\tt i+1}}).$  
	We need to show that ${\tt dim}({\tt ker}\mathcal{M}_{{\tt i+2}})={\tt r}$. Let the columns of the matrix $N\in\R^{\tt im\times \tt r}$ form a basis for ${\tt ker}\mathcal{M}_{{\tt i}}$. By our assumption and Statement $1$ of this lemma it follows that the columns of the matrix $\left[\begin{smallmatrix}
	N\\0_{\tt m,r}
	\end{smallmatrix}\right]$ form a basis for ${\tt ker}\mathcal{M}_{{\tt i+1}}$. Now, to the contrary, assume that ${\tt dim}({\tt ker}\mathcal{M}_{{\tt i+2}})>{\tt r}$. Thus, there exists a vector $w=\left[\begin{smallmatrix}
	w_1\\w_2\\w_3
	\end{smallmatrix}\right]\in{\tt ker}\mathcal{M}_{{\tt i+2}}$, with $w_2\in\R^{\tt im},w_1,w_3\in\Rm$ such that $w\notin{\tt img}\left[\begin{smallmatrix}
	N\\
	0_{\tt m,r}\\
	0_{\tt m,r}
	\end{smallmatrix}\right]$. Since $\left[\begin{smallmatrix}
	w_1\\w_2\\w_3
	\end{smallmatrix}\right]\in{\tt ker}\mathcal{M}_{{\tt i+2}}$, using equation \eqref{eqn:part_markov_matrix} we have
	\begin{eqnarray}{\label{eqn:mark_lemma_1}}
	\mathcal{M}_{{\tt i+2}}\begin{bmatrix}
	w_1\\w_2\\w_3
	\end{bmatrix}=\begin{bmatrix}
	0 & \bD\\\mathcal{M}_{{\tt i+1}} & m_{\tt i+2}
	\end{bmatrix}
	\begin{bmatrix}
	w_1\\w_2\\w_3
	\end{bmatrix}=\begin{bmatrix}
	0 & 0 & \bD\\0 & \mathcal{M}_{{\tt i}} & m_{\tt i+1}\\\bD & \ell_{\tt i+1} & \bC\bA^{\tt i}\bB
	\end{bmatrix}
	\begin{bmatrix}
	w_1\\w_2\\w_3
	\end{bmatrix}=0.
	\end{eqnarray}
	From equation \eqref{eqn:mark_lemma_1}, it follows that $\left[\begin{matrix}
	0 & \bD\\\mathcal{M}_{{\tt i}} & m_{\tt i+1}
	\end{matrix}\right]
	\left[\begin{matrix}
	w_2\\w_3
	\end{matrix}\right] = 
	\mathcal{M}_{{\tt i+1}}\left[\begin{matrix}
	w_2\\w_3
	\end{matrix}\right]=0.$ 
	Thus, $\left[\begin{smallmatrix}
	w_2\\w_3
	\end{smallmatrix}\right]\in{\tt img}\left[\begin{smallmatrix}
	N\\0_{\tt m,r}
	\end{smallmatrix}\right],$ which implies that $w_3=0$. So, from equation \eqref{eqn:mark_lemma_1} we further get that $\left[\begin{smallmatrix}
	0 & \mathcal{M}_{{\tt i}}\\
	\bD & \ell_{\tt i+1}
	\end{smallmatrix}\right]\left[\begin{smallmatrix}
	w_1\\w_2
	\end{smallmatrix}\right]=\mathcal{M}_{{\tt i+1}}\left[\begin{smallmatrix}
	w_1\\w_2
	\end{smallmatrix}\right]=0,$ which in turn implies that $\left[\begin{smallmatrix}
	w_1\\w_2
	\end{smallmatrix}\right]\in{\tt img}\left[\begin{smallmatrix}
	N\\0_{\tt m,r}
	\end{smallmatrix}\right]$. Therefore, $\left[\begin{smallmatrix}
	w_1\\w_2\\w_3
	\end{smallmatrix}\right]\in{\tt img}\left[\begin{smallmatrix}
	N\\0_{\tt m,r}\\0_{\tt m,r}
	\end{smallmatrix}\right]$. But, this is a contradiction. Therefore, our assumption that ${\tt dim}({\tt ker}\mathcal{M}_{{\tt i+2}})>{\tt r}$ cannot be true. Hence, ${\tt dim}({\tt ker}\mathcal{M}_{{\tt i+2}})={\tt r}.$ \\
	\textbf{Inductive step:} Say, ${\tt dim}({\tt ker}\mathcal{M}_{{\tt i+q}})\!\!=\!\!{\tt r}$, for some ${\tt q}\in \Nat$. We need to show that ${\tt dim}({\tt ker}\mathcal{M}_{{\tt i+q+1}}) ={\tt r}$. Using Statement $1$ and Statement $2$ of this lemma, we infer that ${\tt dim}({\tt ker}\mathcal{M}_{{\tt i+q-1}})={\tt r};$
	and the columns of the matrices $\widetilde{N}:=\left[\begin{smallmatrix}
	N\\0_{\tt (q-1)m,r}
	\end{smallmatrix}\right]$ and $\left[\begin{smallmatrix}
	\widetilde{N}\\0_{\tt m,r}
	\end{smallmatrix}\right]$ form the bases for ${\tt ker}\mathcal{M}_{{\tt i+q-1}}$ and ${\tt ker}\mathcal{M}_{{\tt i+q}}$, respectively. Thus, using similar line of arguments as in the base case, we can infer that ${\tt dim}({\tt ker}\mathcal{M}_{{\tt i+q+1}})={\tt r}.$ This completes the proof.
\end{proof1}
The following lemma tells us that, for a system, if $\delta^{(k)}$ is admissible in the input (see Definition \ref{LQRdef:ad_imp_inp}), then $\delta^{(k-1)}$, too, is admissible in the input.
\begin{lemma}\label{lem:no_jump_in_impulses}
	Let $\delta^{(k)}$ is admissible for the system $\Sigma$ defined in equation \eqref{eq:sys_sig}, that is, there exist $u_{i}\!\in\!\Rm,i\!\in\!\{0,1,\dots,k\}$,\! $u_{k}\!\neq\! 0$ such that $u(t)\!:=\!\!\sum\limits_{i=0}^{k}\!u_{i}\delta^{(i)}\!\in\!\mathcal{U_{\tt imp}}$. Then, $\delta^{(k-1)}$ is also admissible for $\Sigma$. In particular, $\tilde{u}(t)\!:=\!\sum\limits_{i=1}^{k}u_{i}\delta^{(i-1)}\!\in\!\Uimp.$
\end{lemma}	
\begin{proof1} Since $u(t)\in\Uimp$, from Proposition \ref{LQRprp:ad_imp_inp} it follows that $G(s)U(s)$ is strictly proper (see Remark \ref{rem:proper_matrices}), where
	\vspace{-.3cm}
	\begin{eqnarray}\label{eqn:transfer_matrix_Gpm}
	G(s)\!:=\!\bC(sI_{\tt n}\!\!-\!\!\bA)^{-1}\!\bB\!+\!\bD \mbox{ is the transfer matrix of $\Sigma$ and } U(s)\!\!:=\!\!\sum\limits_{i=0}^{k}\!\!u_{i}s^{i}.
	\end{eqnarray}  
	Therefore, $\lim\limits_{s\to\infty}G(s)U(s)=0\Leftrightarrow\lim\limits_{s\to\infty}sG(s)\widetilde{U}(s)+\lim\limits_{s\to\infty}G(s)u_0=0,$	where\\ $\widetilde{U}(s):=\sum\limits_{i=1}^{k}u_{i}s^{(i-1)}$. But, from the construction of $G(s)$, it is clear that $G(s)$ is proper. Thus, $w:=\lim\limits_{s\to\infty}G(s)u_0$ is a constant vector. So, $\lim\limits_{s\to\infty}sG(s)\widetilde{U}(s)=-w$, which further implies that $G(s)\widetilde{U}(s)$ is strictly proper. Hence, $\tilde{u}(t)=\sum\limits_{i=1}^{k}u_{i}\delta^{(i-1)}\in\Uimp$. But, $u_k\neq 0$. Therefore, $\delta^{(k-1)}$ is admissible.
\end{proof1}
By repeated application of Lemma \ref{lem:no_jump_in_impulses} it is easy to see that if $u(t)\in\Uimp$, then $\hat{u}(t)=\sum\limits_{i=l}^{k}u_{i}\delta^{(i-l)}\in\Uimp$ for all $l\in\{1,2,\dots,k\}$. So, if $\del{k}$ is admissible in the input, then $\delta,\del{1},\dots,\del{k-1}$ are admissible, too. Thus, if ${\tt dim}(\Uimp)=\f$, then $\delta^{(\f)}$ can not be admissible. This, along with Lemma \ref{lem:markov_mat_prop} bring us to a position to prove Lemma \ref{lem:ad_imp_dimention}.\\

\begin{proof1}{\em of Lemma \ref{lem:ad_imp_dimention}:}
\textbf{1.} We show that ${\rm (a)}\Rightarrow{\rm (b)}\Rightarrow{\rm (c)}\Rightarrow{\rm (a)}.$\\
\textbf{(a)$\Rightarrow$(b):} Suppose ${\tt dim}(\Uimp)=\f$, and $n_{0,\tt i}\delta +n_{1,\tt i}\del{1}+\dots+n_{\tt f-1, i}\del{\f-1}$, ${\tt i}\in\{0,1,\dots,\f-1\}$ be a basis for $\Uimp$, where $n_{\tt{k},\tt i}\in\Rm$, ${\tt k}\in\{0,1,\dots,\f-1\}$. Define the matrix
\begin{eqnarray}{\label{eq:kernel_markov}}
\hN:=\begin{bmatrix}
\hN_0\\\hN_1\\\vdots\\\hN_{\f-1}
\end{bmatrix}:=\begin{bmatrix}
n_{0,0} & n_{0,1} & \dots & n_{0,\f-1}\\
n_{1,0} & n_{1,1} & \dots & n_{1,\f-1}\\
\vdots & \vdots & \ddots & \vdots\\
n_{\f-1,0} & n_{\f-1,1} & \dots & n_{\f-1,\f-1}
\end{bmatrix}\in\R^{\tt fm\times \f}, 
\end{eqnarray}
where $\hN_{\tt i}\in\R^{\tt m\times \f}$, ${\tt i}\in\{0,1,\dots,\f-1\}$. Clearly, $\hN$ is full column-rank and $\hN_{0} v\delta +\hN_{1}v\del{1}+\dots+\hN_{\f-1}v\del{\f-1}\in\Uimp$ for all $v\in\R^{\f}$. Now, for any input from $\Uimp$, the corresponding output of the system $\Sigma$ must be regular. Thus, from Proposition \ref{LQRprp:ad_imp_inp} we have that $G(s)(\hN_0v+\hN_1vs+\dots+\hN_{\f-1}vs^{\f-1})$ is strictly proper for all $v\in\R^{\f}$, where $G(s)$ is as defined in equation \eqref{eqn:transfer_matrix_Gpm}. Next, expressing $G$ by Taylor's series expansion around $s=\infty$, we get
\begin{eqnarray}{\label{eqn:regular_output}}
&\{\bC(sI_{\tt n}&-\bA)^{-1}\bB + \bD\}(\hN_0v + \hN_1vs + \dots + \hN_{\f-1}vs^{\f-1}) =(\bD+\frac{\bC\bB}{s} + \frac{\bC\bA\bB}{s^2} + \dots)(\hN_0v + \hN_1vs + \dots + \hN_{\f-1}vs^{\f-1})\nonumber\\
&&\!\!\!\!\!\!\!\!=s^{\f-1}\bD\hN_{\f-1}v + s^{\f-2}(\bD\hN_{\f-2}v + \bC\bB\hN_{\f-1}v) + s^{\f-3}(\bD\hN_{\f-3}v+\bC\bB\hN_{\f-2}v + \bC\bA\bB\hN_{\f-1}v) + \cdots \nonumber\\
&&\!\!\!\!\!\!\!\!+s(\bD\hN_{1}v+\bC\bB\hN_{2}v + \bC\bA\bB\hN_{3}v + \dots + \bC\bA^{\f-3}\bB\hN_{\f-1}v) + (\bD\hN_{0}v+\bC\bB\hN_{1}v + \bC\bA\bB\hN_{2}v + \dots + \bC\bA^{\f-2}\bB\hN_{\f-1}v)\nonumber \\ 
&&\!\!\!\!\!\!\!\! +  \frac{1}{s}(\bC\bB\hN_{0}v + \bC\bA\bB\hN_{1}v + \dots + \bC\bA^{\f-1}\bB\hN_{\f-1}v) + \dots.
\end{eqnarray}
Since $G(s)(\hN_0v+\hN_1vs+\dots+\hN_{\f-1}vs^{\f-1})$ is strictly proper, the coefficients of $s^0,s^1,\dots,s^{\f-1}$ must all be zero. Therefore, from equation \eqref{eqn:regular_output} it follows that
\begin{eqnarray}{\label{eqn:co-eff_matrix}}
\begin{bmatrix}
0 & 0 & \dots & 0 & \bD\\
0 & 0 & \dots & \bD & \bC\bB\\
0 & 0 & \dots & \bC\bB & \bC\bA\bB\\
\vdots & \vdots & \ddots & \vdots & \vdots\\
\bD & \bC\bB & \dots & \bC\bA^{\f-3}\bB & \bC\bA^{\f-2}\bB
\end{bmatrix}\begin{bmatrix}
\hN_0\\\hN_1\\\vdots\\\hN_{\f-1}
\end{bmatrix}v=\mathcal{M}_{{\tt f}}\hN v=0,
\end{eqnarray}
where $\M{\f}$ is as defined in equation \eqref{eqn:markov_para_matrix}. Equation \eqref{eqn:co-eff_matrix} holds for all $v\in\R^{\f}$; therefore, $\mathcal{M}_{{\tt f}}\hN=0.$ 
Also, $\hN\in\R^{\tt fm\times \f}$ is full column-rank. Hence, ${\tt dim}({\tt ker}\mathcal{M}_{{\tt f}})\geqslant \f$. 

Now, to the contrary, assume that ${\tt dim}({\tt ker}\mathcal{M}_{\f})\neq \f$, then there exists a vector 
${\tt col}(n_{0,\f},n_{1,\f},\dots,n_{\f-1,\f})$ with $n_{\tt{k},\f}\in\Rm$, ${\tt k}\in\{0,1,\dots,\f-1\}$ such that ${\tt col}(n_{0,\f},n_{1,\f},\dots,n_{\f-1,\f})\in{\tt ker}\mathcal{M}_{\f}$, but ${\tt col}(n_{0,\f},n_{1,\f},\dots,n_{\f-1,\f})\notin {\tt img}\hN.$ Since ${\tt col}(n_{0,\f},n_{1,\f},\dots,n_{\f-1,\f})\in{\tt ker}\mathcal{M}_{\f}$, from similar constructions as in equation \eqref{eqn:regular_output} and equation \eqref{eqn:co-eff_matrix}, it follows that $G(s)(n_{0,\f}+n_{1,\f}s+\dots+n_{\f-1,\f}s^{\f-1})$ is strictly proper. Thus, $r(t):=n_{0,\f}\delta+n_{1,\f}\del{1}+\dots+n_{\f-1,\f}\del{\f-1}\in \Uimp$. But, this is a contradiction, because ${\tt col}(n_{0,\f},n_{1,\f},\dots,n_{\f-1,\f})\notin {\tt img}\hN$ and thus $r(t)$ does not belong to the space of inputs spanned by $\{n_{0,\tt i}\delta+n_{1,\tt i}\del{1}+\dots+n_{\f-1,\tt i}\del{\f-1} \}_{{\tt i}\in\{0,1,\dots,\f-1\}}$. Hence, ${\tt dim}({\tt ker}\mathcal{M}_{\f})=\f$.

\noindent Now, we prove that ${\tt dim}({\tt ker}\mathcal{M}_{\f+1})\!=\!\f$. So, to the contrary, assume that ${\tt dim}({\tt ker}\mathcal{M}_{\f+1})\neq\f$. Thus, from Statement $1$ and Statement $2$ of Lemma \ref{lem:markov_mat_prop}, it is evident that ${\tt img}\left[\begin{smallmatrix}
\hN\\0_{\tt m,f}
\end{smallmatrix}\right]\!\!\subseteq\!{\tt ker}\mathcal{M}_{\f+1}$ and ${\tt dim}({\tt ker}\mathcal{M}_{\f+1})>\f$. So, there exists $\hat{v}\!:=\!{\tt col}(v_0,v_1, \dots
, v_{\f})$ with $v_{j}\in\Rm$, $j\in\{0,1,\dots,\f\}$ such that $\hat{v}\in{\tt ker}\mathcal{M}_{\f+1}$, but $\hat{v}\notin{\tt img}\left[\begin{smallmatrix}
\hN\\0_{\tt mf,f}
\end{smallmatrix}\right]$. Using similar method as before, it can be shown that $G(s)(v_0+v_1s +\dots+ v_{\f}s^{\f})$ is strictly proper; thus, $v_0\delta+v_1\del{1} +\dots+ v_{\f}\del{\f}\in\Uimp.$ But, this is a contradiction, because $\hat{v}\!\notin\!{\tt img}\left[\begin{smallmatrix}
\hN\\0_{\tt mf,f}
\end{smallmatrix}\right]$, and hence $v_0\delta+v_1\del{1} +\dots+ v_{\f}\del{\f}\notin{\tt span}\{n_{0,\tt i}\delta+n_{1,\tt i}\del{1}+\dots+n_{\f-1,\tt i}\del{\f-1} \}_{{\tt i}\in\{0,1,\dots,\f-1\}}$. So, ${\tt dim}({\tt ker}\mathcal{M}_{\f+1})=\f$. Hence, Statement (a) implies Statement (b).\\
\textbf{(b)$\Rightarrow$(c):} Suppose ${\tt dim}({\tt ker}\mathcal{M}_{\f})={\tt dim}({\tt ker}\mathcal{M}_{\f+1})=\f$. Since $\mathcal{M}_{1}=\bD$ and ${\tt dim}({\tt ker}D)={\tt d}$, it is clear that ${\tt dim}({\tt ker}\mathcal{M}_{1})={\tt d}$. Now, from Statement $2$ and Statement $3$ of Lemma \ref{lem:markov_mat_prop}, it follows that, for $j\geqslant 1$,
\begin{equation}\label{eqn:b_implies_c}
{\tt dim}({\tt ker}\mathcal{M}_{j+1})-{\tt dim}({\tt ker}\mathcal{M}_{j})\begin{cases}
&\geqslant 1 \mbox{ if } {\tt dim}({\tt ker}\mathcal{M}_{j})<\f\\
&=0 \mbox{ if } {\tt dim}({\tt ker}\mathcal{M}_{j})=\f.
\end{cases}
\end{equation}
If ${\tt dim}(\M{k+1})<\f$, then by repeated application of equation \eqref{eqn:b_implies_c}, we infer that
\begin{equation}\label{eqn:b_implies_c_2}
{\tt dim}(\M{k+1})-{\tt dim}(\M{1})\geqslant (k+1)-1=k.
\end{equation}
Now, to the contrary, assume that ${\tt dim}(\M{\f-\dd+1})<\f$. (${\tt dim}(\M{\f-\dd+1})\ngtr\f$ due to Statement 3 of Lemma \ref{lem:markov_mat_prop}, because we have assumed that ${\tt dim}({\tt ker}\mathcal{M}_{\f})\!\!=\!\!{\tt dim}({\tt ker}\mathcal{M}_{\f+1})\!=\!\f$.) Therefore, substituting $k\!=\!\f\!-\!\dd$ in equation \eqref{eqn:b_implies_c_2}, we have that ${\tt dim}(\M{\f-\dd+1})-{\tt dim}(\M{1})\!\geqslant\!\f\!-\!\dd$. But, ${\tt dim}(\M{1})\!=\!\dd$. So, ${\tt dim}(\M{\f-\dd+1})\geqslant \f$. This is a contradiction to the assumption that ${\tt dim}(\M{\f-\dd+1})<\f$. Therefore, ${\tt dim}(\M{\f-\dd+1})=\f$. Applying equation \eqref{eqn:b_implies_c}, we further infer that ${\tt dim}(\M{\f-\dd+1})={\tt dim}(\M{\f-\dd+2})=\f$. Hence, Statement (b) implies Statement (c).\\
\textbf{(c)$\Rightarrow$(a):} Suppose ${\tt dim}({\tt ker}\mathcal{M}_{\f-\dd+1})\!=\!{\tt dim}({\tt ker}\mathcal{M}_{\f-\dd+2})=\f$ and the columns of $N={\tt col}(N_0,N_1,\dots,N_{\f-\dd})$ with $N_{i}\in\R^{\m\times\f}, i\in\{0,1,\dots,\f-\dd\}$ form a basis for ${\tt ker}\mathcal{M}_{\f-\dd+1}$. Thus, from Lemma \ref{lem:markov_mat_prop} it follows that the dimension of ${\tt ker}\mathcal{M}_{{\tt f-d+1+q}}=\f$ for all ${\tt q}\in\Nat$ and a basis for ${\tt ker}\mathcal{M}_{{\tt f-d+1+q}}$ is the columns of $\left[\begin{smallmatrix}
N\\0_{\tt qm,f}
\end{smallmatrix}\right]$. Since ${\tt img}N={\tt ker}\mathcal{M}_{\f-\dd+1}$, by similar constructions as in equation \eqref{eqn:regular_output} and equation \eqref{eqn:co-eff_matrix}, it follows that $G(s)(N_0v+N_1vs+\dots+N_{\f-\dd}vs^{\f-\dd})$ is strictly proper for all $v\in\R^{\f}$. Consequently, $N_0v\delta+N_1v\del{1}+\dots+N_{\f-\dd}v\del{\f-\dd}\in\Uimp$ for all $v\in\R^{\f}$. So, from the fact that ${\tt dim}({\tt img}~{\tt col}(N_0,N_1,\dots,N_{\f-\dd}))=\f$, we further have that ${\tt dim}(\Uimp)\geqslant \f$. Now, to the contrary, we assume that ${\tt dim}(\Uimp)> \f$. So, there exists an input $\tilde{r}(t):=n_{0}\delta+n_{1}\del{1}+\dots+n_{\f-\dd+\tt q}\del{\f-\dd+\tt q}\in\Uimp$ with $n_{j}\in\Rm, j\in\{0,1,\dots,\f-\dd+{\tt q}\}$ for some ${\tt q}\in \Nat$ such that ${\tt col}(n_{0},n_{1},\dots,n_{\f-\dd+\tt q})\notin{\tt img}\left[\begin{smallmatrix}
N\\0_{\tt qm,f}
\end{smallmatrix}\right]$. But, since $\tilde{r}(t)\in\Uimp$, we must have that $G(s)(n_{0}+n_{1}s+\dots+n_{\f-\dd+\tt q}s^{(\f-\dd+\tt q)})$ is strictly proper. Next, by similar constructions as in equation \eqref{eqn:regular_output} and equation \eqref{eqn:co-eff_matrix} is evident that ${\tt col}(n_{0},n_{1},\dots,n_{\f-\dd+\tt q})\in{\tt ker}\mathcal{M}_{{\tt f-\dd+1+q}}$; which, in turn, implies that ${\tt col}(n_{0},n_{1},\dots,n_{\f-\dd+\tt q})\in{\tt img}\left[\begin{smallmatrix}
N\\0_{\tt qm,f}
\end{smallmatrix}\right]$. This is a contradiction. Hence, ${\tt dim}(\Uimp)=\f$. Thus, Statement (c) implies Statement (a). This completes the proof of Statement 1.\\

\noindent \textbf{2:} \textbf{$\Uimp\subseteq\Delta$:} Suppose $u_0\delta +u_1\del{1}+\dots+u_{\f-\tt d}\delta^{(\f-\tt d)}+\dots+u_{\tt q-1}\del{\tt q-1}\in\mathcal{U}_{\tt imp}$ is arbitrary, where $u_{i}\in\Rm, i\in\{0,1,\dots,{\tt q}-1\}$ for some ${\tt q}\in \Nat$, ${\tt q}>\f-\dd$. Then, by Proposition \ref{LQRprp:ad_imp_inp}, $G(s)(u_0 +u_1s+\dots+u_{\f-\tt d}s^{(\f-\tt d)}+\dots+\dots+u_{\tt q-1}s^{(\tt q-1)})$ is strictly proper. So, it is easy to verify that $
\hat{u}:={\tt col}(u_0,\dots,u_{\f-\tt d},\dots,u_{\tt q-1})$ $\in{\tt ker}\mathcal{M}_{{\tt q}}$. But, ${\tt dim}(\Uimp)=\f$. So, from Statement $3$ of Lemma \ref{lem:markov_mat_prop}, and Lemma \ref{lem:ad_imp_dimention}, it can be inferred that 
$$\nonumber {\tt dim}({\tt ker}\mathcal{M}_{{\tt f-d+1}})={\tt dim}({\tt ker}\mathcal{M}_{{\tt f-d+2}})=\f={\tt dim}({\tt ker}\mathcal{M}_{{\tt q}}).$$
Now, if the columns of $N\in\R^{\tt m(f-d+1)\times \f}$ form a basis for ${\tt ker}\mathcal{M}_{{\tt f-d+1}}$, then from Statement $1$ of Lemma \ref{lem:markov_mat_prop} we get that ${\tt ker}\mathcal{M}_{{\tt q}}={\tt img}\left[\begin{smallmatrix}
N\\0
\end{smallmatrix}\right]\in\R^{\tt mq\times f}$. Thus, $\hat{u}\in{\tt img}\left[\begin{smallmatrix}
N\\0
\end{smallmatrix}\right]$. So, ${\tt col}(u_0,u_1,\dots,u_{\f-\tt d})\in{\tt img}N={\tt ker}\mathcal{M}_{{\tt f-d+1}}$, and $u_{\tt f-d+1}=u_{\tt f-d+2}=\dots=u_{\tt q-1}=0$. Hence, $\Uimp\subseteq\Delta$.

\noindent \textbf{$\Delta\subseteq\Uimp$:} Let ${\tt col}(u_0,u_1,\dots,u_{\f-\tt d})\in{\tt ker}\mathcal{M}_{{\tt f-d+1}}$. 
Since ${\tt dim}(\Uimp)=\f$, from Lemma \ref{lem:ad_imp_dimention} it follows that ${\tt dim}({\tt ker}\mathcal{M}_{{\tt f-d+1}})=\f$. Assume that the columns of ${\tt col}(N_0,N_1,\dots,N_{\f-\tt d})$ form a basis for ${\tt ker}\mathcal{M}_{{\tt f-d+1}}$, where $N_{i}\in\R^{\m\times\f},i\in\{0,1,\dots,\f-\dd\}$. Then, there exists $v\in\R^{\f}$ such that $\left[\begin{smallmatrix}
u_0\\u_1\\\vdots\\u_{\f-\tt d}
\end{smallmatrix}\right]=\left[\begin{smallmatrix}
N_0\\N_1\\\vdots\\N_{\f-\tt d}
\end{smallmatrix}\right]v.$ Thus, by using similar constructions as in equation \eqref{eqn:regular_output} and equation \eqref{eqn:co-eff_matrix}, it is evident that $G(s)(u_0 +u_1s+\dots+u_{\f-\tt d}s^{(\f-\tt d)})=G(s)(N_0v +N_1vs+\dots+N_{\f-\tt d}vs^{(\f-\tt d)})$ is strictly proper; this further implies that $u_0\delta +u_1\del{1}+\dots+u_{\f-\tt d}\delta^{(\f-\tt d)}\in\Uimp$. Hence, $\Delta\subseteq\Uimp$. This completes the proof.	
\end{proof1}

Lemma \ref{lem:ad_imp_dimention} and Lemma \ref{lem:markov_mat_prop} provide us with the necessary tools to prove Theorem \ref{LQRthm:fast_space}) which characterizes the strongly reachable subspace $\Rea$ .\\

\begin{proof1}{\em of Theorem \ref{LQRthm:fast_space}:}
	\textbf{1.} From equation \ref{LQReqn:fast_recursion}, we know that the recursive algorithm for computing $\Rea$ is given by
	\begin{align*}
	\mathcal{R}_0 := \{0\} \subsetneq \R^{\n}, \mbox{ and } \mathcal{R}_{i+1} := 
	\begin{bmatrix} \bA & \bB\end{bmatrix}\left\{(\mathscr{W}_{i} \oplus \mathscr{P}) \cap {\tt ker}
	\begin{bmatrix}\bC & \bD\end{bmatrix}\right\} \subseteq \Rea, 
	\end{align*}	
	where $\mathscr{W}_{i} := \left\{\left[\begin{smallmatrix}w\\0\end{smallmatrix}\right] \in \R^{{\n}+{\m}}\,|\, 
	w \in \mathcal{R}_{i}\right\}$ and $\mathscr{P} := 
	\left\{\left[\begin{smallmatrix}0\\\alpha\end{smallmatrix}\right] \in \R^{{\n}+{\m}}\,|\, 
	\alpha \in \Rm\right\}$. If $\mathcal{R}_{i+1}=\mathcal{R}_{i}$ for some $i\in\Nat$, then $\Rea=\mathcal{R}_{i}$.\\
	Notice that $\mathscr{W}_{i}\oplus\mathscr{P}:= \left\{\left[\begin{smallmatrix}w\\\alpha\end{smallmatrix}\right] \in \R^{{\n}+{\m}}\,|\, 
	w \in \mathcal{R}_{i} \mbox{ and }\alpha\in\Rm\right\}.$ Thus,
	\begin{align}\label{eqn:mathscr_Ui}
	\mathscr{U}_{i}\!:=\!(\mathscr{W}_{i}\! \oplus\! \mathscr{P}) \cap {\tt ker}
	\left[\begin{smallmatrix}C & D\end{smallmatrix}\right]\!=\!\left\{\left[\begin{smallmatrix}w\\\alpha\end{smallmatrix}\right] \in \R^{{\n}+{\m}}~|~\left[\begin{smallmatrix}C & D\end{smallmatrix}\right]\left[\begin{smallmatrix}
	w\\\alpha
	\end{smallmatrix}\right]=0 \mbox{ and } w \in \mathcal{R}_{i}\right\}.
	\end{align}
	Therefore, $\mathcal{R}_{i+1}$ can be rewritten as
    \begin{align}\label{eqn:iteration_new_form}
     \mathcal{R}_{i+1}=\begin{bmatrix}
     A & B
     \end{bmatrix}(\mathscr{U}_{i}), \mbox{ where $\mathscr{U}_{i}$ is defined in equation \eqref{eqn:mathscr_Ui}.}
    \end{align}
	\textbf{Claim:} $\mathcal{R}_{i}$ is given by $\mathcal{R}_{i}={\tt img}\left[\begin{smallmatrix}
	\bB & \bA\bB & \dots & \bA^{i-1}\bB
	\end{smallmatrix}\right]\hN_{i}$, where the columns of $\hN_{i}$ form a basis for ${\tt ker}\mathcal{M}_{{i}}$ (defined in equation \eqref{eqn:markov_para_matrix}).\\
    We prove this claim by induction.\\
	\underline{Base case ({i}=1):} Since $\mathcal{R}_0\!=\!\{0\}\subsetneq \Rn$, we have $\mathscr{U}_0\!=\!\left\{\left[\begin{smallmatrix}0\\\alpha\end{smallmatrix}\right] \in \R^{{\n}+{\m}}~|~D\alpha\!=\!0\right\}.$ So,
	$$
	\mathcal{R}_1=\left[\begin{smallmatrix}
	\bA & \bB
	\end{smallmatrix}\right]\left\{\left[\begin{smallmatrix}0\\\alpha\end{smallmatrix}\right] \in \R^{{\n}+{\m}}~|~D\alpha\!=\!0\right\}=B({\tt ker}D).
	$$
    Thus, $\mathcal{R}_1={\tt img}B\hN_1, \mbox{ where the columns of }
	\hN_1 \mbox{ form a basis for } {\tt ker}\bD = {\tt ker}\mathcal{M}_{1}$ (since $\M{1}=D$). This proves the base case.\\
	\underline{Inductive step:} We assume that $\mathcal{R}_{i}={\tt img}\left[\begin{smallmatrix}
	\bB & \bA\bB & \dots & \bA^{i-1}\bB
	\end{smallmatrix}\right]\hN_{i}$, where the columns of $\hN_{i}$ form a basis for the kernel of $\mathcal{M}_{{i}}$. 
	We need to show that
	$$\mathcal{R}_{i+1}={\tt img}\left[\begin{smallmatrix}
	\bB & \bA\bB & \dots & \bA^{i}\bB
	\end{smallmatrix}\right]\hN_{i+1},$$
	where the columns of $\hN_{i+1}$ form a basis for ${\ker}\mathcal{M}_{{i+1}}$. From equation \eqref{eqn:mathscr_Ui}, we have that $\mathscr{U}_{i}\!=\!\left\{\left[\begin{smallmatrix}w\\\alpha\end{smallmatrix}\right] \in \R^{{\n}+{\m}}~|~\left[\begin{smallmatrix}C & D\end{smallmatrix}\right]\left[\begin{smallmatrix}
	w\\\alpha
	\end{smallmatrix}\right]\!=\!0 \mbox{ and } w \!\in\! \mathcal{R}_{i}\right\}.$ But, by the inductive hypothesis, $\mathcal{R}_{i}\!=\!\left[\begin{smallmatrix}
	\!\bB & \bA\bB & \dots & \bA^{i-1}\bB
	\end{smallmatrix}\!\right]\!({\tt ker}\M{i}).$ Thus, $\mathscr{U}_{i}$ can be rewritten as
	\begin{align*}
	&\mathscr{U}_{i}\!=\!\left\{\left[\begin{smallmatrix}w\\\alpha\end{smallmatrix}\right] \!\in\! \R^{{\n}+{\m}}~|~ w\!=\!\left[\begin{smallmatrix}
	\bB & \bA\bB & \dots & \bA^{i-1}\bB
	\end{smallmatrix}\right]v \mbox{ and } Cw\!+\!D\alpha=0, \mbox{ where } v \in {\tt ker}\M{i}\right\}\\
	&=\!\left\{\left[\begin{smallmatrix}w\\\alpha\end{smallmatrix}\right] \in \R^{{\n}+{\m}}~|~w\!=\!\left[\begin{smallmatrix}
	\bB & \bA\bB & \dots & \bA^{i-1}\bB
	\end{smallmatrix}\right]v \mbox{ and }\ell_{i+1}v\!+\!D\alpha\!=\!0, \mbox{ where } v \in {\tt ker}\M{i}\right\}\\
	&=\!\left\{\left[\begin{smallmatrix}w\\\alpha\end{smallmatrix}\right] \in \R^{{\n}+{\m}}~|~w\!=\!\left[\begin{smallmatrix}
	\bB & \bA\bB & \dots & \bA^{i-1}\bB
	\end{smallmatrix}\right]v \mbox{ and }\left[\begin{smallmatrix}
	\M{i} & 0\\
	\ell_{i+1} & D
	\end{smallmatrix}\right]\left[\begin{smallmatrix}
	v\\\alpha
	\end{smallmatrix}\right]\!=\!0 \mbox{ for some }v\in\R^{i\m}\right\},
	\end{align*}
	where $\ell_{i+1}:=\left[\begin{smallmatrix}
	CB & CAB & \dots & CA^{i-1}B
	\end{smallmatrix}\right].$ Therefore,
	\begin{align*}
	\mathcal{R}_{i+1}&=\mathcal{R}_{i+1}\!\!=\!\! \left[\begin{smallmatrix}
	A & B
	\end{smallmatrix}\right](\mathscr{U}_{i})\!=\!\left\{A\left[\begin{smallmatrix}
	B & AB & \dots & A^{i-1}B
	\end{smallmatrix}\right]v+B\alpha\!~\vert~\! \left[\begin{smallmatrix}
	\M{i} & 0\\
	\ell_{i+1} & D
	\end{smallmatrix}\right]\!\!\left[\begin{smallmatrix}
	v\\\alpha
	\end{smallmatrix}\right]\!=\!0 \right\}\\
	&=\left\{B\alpha+\left[\begin{smallmatrix}
	AB & A^2B & \dots & A^{i}B
	\end{smallmatrix}\right]v\!~\vert~\! \left[\begin{smallmatrix}
	0 & \M{i}\\
	D & \ell_{i+1}
	\end{smallmatrix}\right]\left[\begin{smallmatrix}
	\alpha\\v
	\end{smallmatrix}\right]\!=\!0 \right\}.
	\end{align*}
	But, $\left[\begin{smallmatrix}
	0 & \M{i}\\
	D & \ell_{i+1}
	\end{smallmatrix}\right]=\M{i+1}.$ Hence, $\mathcal{R}_{i+1}={\tt img}\left[\begin{smallmatrix}
		\bB & \bA\bB & \dots & \bA^{i}\bB
	\end{smallmatrix}\right]\hN_{i+1},$ where the columns of $\hN_{i+1}$ form a basis for ${\tt ker}\mathcal{M}_{i+1}$. This proves the claim.\\
	Now, since ${\dim}(\Uimp)=\f$, from Statement $1$ of Lemma \ref{lem:ad_imp_dimention} it follows that ${\tt dim}({\tt ker}\mathcal{M}_{{\tt f-d+1}})={\tt dim}({\tt ker}\mathcal{M}_{{\tt f-d+2}})=\f,$ 
	where ${\tt d}:={\tt dim}({\tt ker}D)$. From Statement $3$ of Lemma \ref{lem:markov_mat_prop}, this further implies that ${\tt dim}({\tt ker}\mathcal{M}_{j})=\f$ for all $j\geqslant\f-{\tt d}+1$. Moreover, it is evident from Lemma \ref{lem:markov_mat_prop} that columns of the matrix $\left[\begin{smallmatrix}
	N\\0_{(j-\f+\dd-1),\f}
	\end{smallmatrix}\right]\in\R^{j\m\times \f}$ form a basis for ${\tt ker}\mathcal{M}_{j}$ for all $j\geqslant\f-{\tt d}+1$, where ${\tt img}N={\tt ker}\mathcal{M}_{{\tt f-d+1}}$. So, using the claim that we have proved, $\mathcal{R}_{\tt f-d+1}=\left[\begin{smallmatrix}
	B & AB & \dots & A^{\tt f-d}B
	\end{smallmatrix}\right]N$. Further, since ${\tt ker}\mathcal{M}_{j}={\tt img}\left[\begin{smallmatrix}
	N\\0_{(j-\f+\dd-1),\f}
	\end{smallmatrix}\right]$ for all $j\geqslant {\tt f-d+1}$, we must have 
	\begin{align*}
	\mathcal{R}_{j} &=\left[\begin{smallmatrix}
	\bB & \bA\bB & \dots & \bA^{\tt f-d}\bB & \dots & \bA^{j-1}\bB
	\end{smallmatrix}\right]\left[\begin{smallmatrix}
	N\\0_{(j-\f+\dd-1),\f}
	\end{smallmatrix}\right]=\left[\begin{smallmatrix}
	\bB & \bA\bB & \dots & \bA^{\tt f-d}\bB
	\end{smallmatrix}\right]N =\mathcal{R}_{\tt f-d+1}.
	\end{align*}
	Hence, $\Rea=\left[\begin{smallmatrix}
	\bB & \bA\bB & \dots & \bA^{\tt f-d}\bB
	\end{smallmatrix}\right]N$. This completes the proof of Statement 1.
	
	\noindent \textbf{2.} From Lemma \ref{lem:ad_imp_dimention}, ${\tt dim}(\Uimp)={\tt dim}({\tt ker}\mathcal{M}_{{\tt f-d+1}})={\tt dim}({\tt ker}\mathcal{M}_{{\tt f-d+2}})=\f.$ Also, it is given that columns of $N\in\R^{\tt m(f-d+1)\times \f}$ form a basis for ${\tt ker}\M{\f-\dd+1}$. Therefore, $N=:{\tt col}(N_0,N_1,\dots,N_{\tt f-d})$ is full column-rank, where $N_{i}\in\R^{\tt m\times \f}, 0\leqslant i\leqslant (\tt f-d)$. Recall that $\Rea$ is given by the image of the matrix $\left[\begin{smallmatrix}
	B & AB & \dots & A^{\tt f-d}B
	\end{smallmatrix}\right]N\in\R^{\n \times \f}$. Thus, ${\tt dim\,}{\Rea}\leqslant\f$. Now, to the contrary, we assume that ${\tt dim}(\Rea)<\f$, which implies that there exists $v\in\R^{\f}\backslash\{0\}$ such that 
	\begin{equation}{\label{eqn:fast_fcr_1}}
	\left[\begin{smallmatrix}
	B & AB & \dots & A^{\tt f-d}B
	\end{smallmatrix}\right]N v=0.
	\end{equation}
	Define $\mathcal{N}:=\left[\begin{smallmatrix}
	C & 0 & \dots & 0\\
	0 & CA & \dots & 0\\
	\vdots & \vdots & \ddots & \vdots\\
	0 & 0 & \dots & CA^{\tt f-d}
	\end{smallmatrix}\right]\left[\begin{smallmatrix}
	B & AB & \dots & A^{\tt f-d}B\\
	B & AB & \dots & A^{\tt f-d}B\\
	\vdots & \vdots & \ddots & \vdots\\
	B & AB & \dots & A^{\tt f-d}B
	\end{smallmatrix}\right].$
	Then, from equation \eqref{eqn:fast_fcr_1}, it follows that $\mathcal{N}N v=0$. Also, it is easy to verify that $\mathcal{M}_{{\tt 2f-2d+2}}=\left[\begin{smallmatrix}
	0 & \mathcal{M}_{{\tt f-d+1}}\\
	\mathcal{M}_{{\tt f-d+1}} & \mathcal{N}
	\end{smallmatrix}\right].$ This implies that $\left[\begin{smallmatrix}
	Nv\\Nv
	\end{smallmatrix}\right]\in{\tt ker}\mathcal{M}_{{\tt 2f-2d+2}}$, because $\mathcal{N}N v=0$ and $\M{\f-\dd+1}N=0$. Next, from Lemma \ref{lem:markov_mat_prop} we conclude that ${\ker}\mathcal{M}_{{\tt 2f-2d+2}}={\tt img}\left[\begin{smallmatrix}
	N\\0
	\end{smallmatrix}\right]\subseteq\R^{\tt m(2f-2d+2)}$; this, in turn, implies that $Nv=0$. But, $N$ is full column-rank $\Rightarrow v=0$. This is a contradiction. So, ${\tt ker}\left[\begin{smallmatrix}
	B & AB & \dots & A^{\tt f-d}B
	\end{smallmatrix}\right]N=\{0\}$, and hence ${\tt dim}(\Rea)=\f$. This completes the proof.
\end{proof1}
The salient points of Theorem \ref{LQRthm:fast_space} are the following: firstly, it shows that the spaces $\Uimp$ and $\Rea$ have the same dimension; and secondly it also shows that if 
$$\Uimp=\left\{\hN_0 v\delta+\hN_1 v\del{1}+\dots+\hN_{\tt f-d} v\del{\tt f-d} ~\vert~ v\in\R^{\f}\right\},$$
$$\text{then }\Rea\!\!=\!\!\left[\begin{matrix}
\bB & \bA\bB & \dots & \bA^{\tt f-d}\bB
\end{matrix}\right]{\tt col}(\hN_0,\hN_1,\dots,\hN_{\tt f-d}).
$$
\subsection{Dimension of the strongly reachable subspace from the transfer matrix}\label{LAAsec:sub_dim_fast_space}
In this section we compute the dimension of the strongly reachable subspace, $\Rea$, from the transfer matrix of a given system. In addition to the results developed in Section \ref{LAAsec:sub_char_fast_space}, we need two more auxiliary results to achieve this task. The first of these results is the following lemma which is obtained by combining two results from \cite{HauSil:83} along with Theorem \ref{thm:slow_space}.
\begin{lemma}\label{lem:square_system_dimension}
	Define the system $\Theta: \ddt x(t)=\tA x(t)+\tB u(t)$ and $y(t)=\tC x(t)+\tD u(t)$, where $\tA\in\R^{\N}, \tB,\tC\in\R^{\N\times{\tt M}},$ and $D\in\R^{{\tt M}\times{\tt M}}$. Define $\widehat{G}(s):=\tC(sI_{\N}-\tA)\tB+\tD\in\R(s)^{{\tt M}\times{\tt M}}$, the transfer matrix of $\Theta$. Assume that $\widehat{G}(s)$ is invertible as a rational function matrix. Define $\Ns:={\tt deg}\{{\tt num}{\tt det}\widehat{G}(s)\}$\footnote{${\tt num}{\tt det}\widehat{G}(s)$ is the numerator of ${\tt det}\widehat{G}(s)$ before the pole-zero cancellation, if any. For example: if ${\tt det}\widehat{G}(s)=\frac{(s+1)(s+2)}{(s+1)(s+3)}$, then ${\tt num}{\tt det}\widehat{G}(s)=(s+1)(s+2)$, not $(s+2)$.}. Then the following are true:
	\begin{enumerate}
		\item[{\rm 1.}] Dimension of the weakly unobservable subspace $(\mathcal{W}_u)$ of $\Theta$ is $\Ns$.
		\item[{\rm 2.}] Dimension of the strongly reachable subspace $(\mathcal{S}_r)$ of $\Theta$ is $(\N-\Ns)$.
	\end{enumerate}
\end{lemma}
\begin{proof1}
	\textbf{1.} Define $U_1\!\!:=\!\!\left[\begin{smallmatrix}
	I_{\N} & 0\\
	0 & 0_{{\tt M,M}}
	\end{smallmatrix}\right]$ and $U_2\!\!:=\!\!\left[\begin{smallmatrix}
	\tA & \tB\\
	\tC & \tD
	\end{smallmatrix}\right]$. Then, by Theorem \ref{thm:slow_space}, ${\tt dim}(\mathcal{W}_u)={\tt deg}\{{\tt det}(sU_1-U_2)\}$. But, notice that ${\tt det}(sU_1-U_2)={\tt det}\{-\tD-\tC(sI_{\N}-\tA)^{-1}\tB\}{\tt det}(sI_{\N}-\tA)=(-1)^{\tt M}{\tt det}\widehat{G}(s){\tt det}(sI_{\N}-\tA)=(-1)^{\tt M}{\tt num}{\tt det}\widehat{G}(s).$ Therefore, ${\tt deg}\{{\tt det}(sU_1-U_2)\}=\Ns$ and hence ${\tt dim}(\mathcal{W}_u)=\Ns$.\\
	\textbf{2.} By \cite[Theorem 3.24]{HauSil:83}, it follows that $\mathcal{W}_u+\mathcal{S}_r=\R^{\N}$. Further, since $\widehat{G}(s)$ is invertible as a rational function matrix, by \cite[Theorem 3.26]{HauSil:83}, $\mathcal{W}_u\cap\mathcal{S}_r=\{0\}$. Thus, $\mathcal{W}_u\oplus\mathcal{S}_r=\R^{\N}$ and hence ${\tt dim}(\mathcal{S}_r)=\N-\Ns.$
\end{proof1}
Notice that the system considered in Lemma \ref{lem:square_system_dimension} is a square system, that is, the input cardinality and the output cardinality of the system are equal. Hence, the proposition is not applicable to systems having non-square transfer matrices. Our aim is to provide a result which is applicable to systems having non-square but left-invertible (as a rational function matrix) transfer matrices. The following auxiliary lemma becomes useful in proving this result.
\begin{lemma}\label{lem:strictly_proper}
Let $r(s)\in\R(s)^{\bullet\times 1}$. Then, $r(s)$ is strictly proper (recall Remark \ref{rem:proper_matrices}) if and only if $r(-s)^Tr(s)$ is strictly proper.
\end{lemma}
\begin{proof1}
	Let $r(s)$ is strictly proper. Then, $\lim\limits_{\omega\to\infty} r(j\omega)=0$, where $j=\sqrt{-1}$ and $\omega\in\R$. Now,
	$$
	r(-j\omega)^Tr(j\omega)=\|r(j\omega)\|_2^2.
	$$
	Thus, $\lim\limits_{\omega\to\infty}\!\!r(-j\omega)^Tr(j\omega)\!=\!\lim\limits_{\omega\to\infty}\!\|r(j\omega)\|_2^2\!=\!\|\!\!\lim\limits_{\omega\to\infty}\!r(j\omega)\|_2^2\!=\!0.$ Hence, $r(-s)^Tr(s)$ is strictly proper. The converse can be proved in a similar manner.
\end{proof1}
Now, we finally prove the main result of this section, which renders the dimension of the strongly reachable subspace ($\Rea$) from the transfer matrix. Before we state this theorem, note that if the transfer matrix $G(s)$ is left-invertible, then $G(-s)^TG(s)$ is square and non-singular. Further, if $\lambda\in\C$ is a root of ${\tt num}{\tt det}G(-s)^TG(s)$, then $-\lambda$, too, is a root. Thus, ${\tt num}{\tt det}G(-s)^TG(s)$ is an even polynomial and consequently has an even degree. 
\begin{theo}\label{thm:dim_strong_reach}
 Consider the system $\Sigma$ and its transfer matrix $G(s)\!\in\!\R(s)^{\p\times\m}$ as defined in equation \eqref{eq:sys_sig} and equation \eqref{eqn:transfer_matrix_Gpm}, respectively. Assume that $G(s)$ is left-invertible as a rational function matrix. Define 
 $$2\ns:={\tt deg}\{{\tt num}{\tt det}G(-s)^TG(s)\}$$
 Then, dimension of the strongly reachable subspace $\Rea$ of $\Sigma$ is $\nf:=\n-\ns$.
\end{theo}
\begin{proof1}
	Since $G(s)\in\R(s)^{\p\times\m}$ is possibly non-square, Lemma \ref{lem:square_system_dimension} is inapplicable to $\Sigma$. The idea of this proof is to construct a square transfer matrix for which the dimension of the strongly reachable subspace will be same as that of $\Sigma$. So, the dimension of $\Rea$ can be computed via this square system by applying Lemma \ref{lem:square_system_dimension}. First, consider the matrices $J(s)\in\R[s]^{\p\times \m}$, and $T(s)\in\R[s]^{\m\times\m}$ such that $G(s)=J(s)T(s)^{-1}$. But, it is given that $G(s)=C(sI_{\n}-A)^{-1}B+D$. From this structure of $G(s)$, it is clear that $G(s)=J(s)T(s)^{-1}$ is proper. Now, $G(-s)^TG(s)=T(-s)^{-T}J(-s)^TJ(s)T(s)^{-1}$ is non-singular, so $J(-s)^TJ(s)$, too, is non-singular. Also, since $J(-j\omega)^TJ(j\omega)>0$ for all $\omega\in\R$, from \cite[Proposition 5.6]{WilTre:98}, it follows that there exists $P(s)\in\R[s]^{\m\times\m}$ Hurwitz such that $J(-s)^TJ(s)=P(-s)^TP(s)$. Thus, $G(-s)^TG(s)=T(-s)^{-T}P(-s)^TP(s)T(s)^{-1}=G_{\tt sq}(-s)^TG_{\tt sq}(s)$, where $G_{\tt sq}(s):=P(s)T(s)^{-1}\in\R(s)^{\m \times\m}$ is non-singular. In what follows, we show that the dimension of the strongly reachable subspace of a system having transfer matrix $G_{\tt sq}$ is same as that of a system having transfer matrix $G(s)$.\\
	Next, recall from Proposition \ref{LQRprp:ad_imp_inp} that $u(t)=\sum\limits_{i=0}^{k}u_{i}\del{i}$ with $u_{i}\in\R^{\m}, 0\leqslant i\leqslant k$, is an admissible impulsive input for $\Sigma$ if and only if $G(s)U(s)$ is strictly proper, where $U(s):=\sum_{i=0}^{k}u_{i}s^{i}$. Now, by Lemma \ref{lem:strictly_proper}, $G(s)U(s)$ is strictly proper $\Leftrightarrow$ $U(-s)^TG(-s)^TG(s)U(s)$ is strictly proper $\Leftrightarrow$ $U(-s)^TG_{\tt sq}(-s)^TG(s)_{\tt sq}U(s)$ is strictly proper (because $G(-s)^TG(s)=G_{\tt sq}(-s)^TG(s)_{\tt sq}$) $\Leftrightarrow$ $G(s)_{\tt sq}U(s)$ is strictly proper. Hence, $u(t)$ is an admissible impulsive input for a system having transfer matrix $G(s)$ if and only if $u(t)$ is an admissible impulsive input for a system having transfer matrix $G_{\tt sq}(s)$. Therefore, by Theorem \ref{LQRthm:fast_space}, it follows that dimensions of the strongly reachable subspaces for the systems having transfer matrix $G(s)$ and the systems having transfer matrix $G_{\tt sq}$ are same.\\ Next, since ${\tt deg}\{{\tt num} {\tt det}G_{\tt sq}(-s)^TG(s)_{\tt sq}\}={\tt deg}\{{\tt num} {\tt det}G(-s)^TG(s)\}=2\ns$, we conclude that ${\tt deg}\{{\tt num} {\tt det}G_{\tt sq}(s)\}=\ns$. Thus, from Statement 2 of Lemma \ref{lem:square_system_dimension}, we get that the dimension of the strongly reachable subspace of a system having transfer matrix $G_{\tt sq}(s)$ is $\n-\ns=\nf$. Hence, the dimension of the strongly reachable subspace of $\Sigma$=${\tt dim}(\Rea)=\nf$.
\end{proof1}

\section{Conclusion}\label{LAAsec:conclusion} In this paper we used the recursive algorithms provided in \cite{HauSil:83} to provide closed-form representations for the weakly unobservable and strongly reachable subspaces. We showed that the dimensions of these spaces can be directly read off from the transfer matrix of the given system. We also characterized the admissible impulsive inputs that guarantees regular output. The explicit relation between the space of admissible impulsive inputs ($\Uimp$) and the strongly reachable subspace ($\Rea$) has been established in this paper. We also showed that the dimension of both the spaces $\Uimp$ and $\Rea$ is the same.
\bibliography{reference}

\begin{thebibliography}{10}
\expandafter\ifx\csname url\endcsname\relax
  \def\url#1{\texttt{#1}}\fi
\expandafter\ifx\csname urlprefix\endcsname\relax\def\urlprefix{URL }\fi
\expandafter\ifx\csname href\endcsname\relax
  \def\href#1#2{#2} \def\path#1{#1}\fi

\bibitem{Wil:80}
J.~Willems, Almost ${A}$mod$({B})$-invariant subspaces, Ast\'erisque 75-76
  (1980) 239--248.

\bibitem{Won:85}
W.~Wonham, Linear {M}ultivariable {C}ontrol: {A} {G}eometric {A}pproach,
  Springer-{V}erlag, New York, 1985.

\bibitem{HauSil:83}
M.~Hautus, L.~Silverman, System structure and singular control, Linear Algebra
  and its Application 50 (1983) 369--402.

\bibitem{HeeSchWei:00}
W.~Heemels, H.~Schumacher, S.~Weiland, Linear complementarity systems, SIAM
  Journal on Applied Mathematics 60~(4) (2000) 1234--1269.

\bibitem{HeeKanSchJoh:11}
W.~Heemels, M.~K. Camlibel, J.~M. Schumacher, B.~Brogliato, Observer-based
  control of linear complementarity systems, International Journal of Robust
  and Nonlinear Control 21~(10) (2011) 1193--1218.

\bibitem{Ter:98}
W.~J. Terrell, An input-output representation for implicit linear time-varying
  systems, Linear algebra and its applications 271~(1-3) (1998) 221--234.

\bibitem{KazNtoPer:18}
C.~Kazantzidou, L.~Ntogramatzidis, T.~Perez, Computation of regular friends for
  output-nulling and reachability subspaces of linear time-invariant descriptor
  systems, in: 2018 European Control Conference (ECC), IEEE, 2018, pp.
  2505--2510.

\bibitem{And:75}
B.~D. Anderson, Output-nulling invariant and controllability subspaces, IFAC
  Proceedings Volumes 8~(1) (1975) 337--345.

\bibitem{Fuh:05}
P.~Fuhrmann, Autonomous subbehaviours and output nulling subspaces,
  International Journal of control 78~(17) (2005) 1378--1411.

\bibitem{PadNto:20}
F.~Padula, L.~Ntogramatzidis, Fixed poles in the disturbance decoupling by
  dynamic output feedback for systems with direct feedthrough matrices,
  Automatica 121 (2020) 109159.

\bibitem{PadFerNto:21}
F.~Padula, A.~Ferrante, L.~Ntogramatzidis, Eigenstructure assignment in linear
  geometric control, Automatica 124 (2021) 109363.

\bibitem{QaiBhaPal:20}
I.~Qais, D.~Pal, C.~Bhawal, A geometric characterization of the slow space of
  the hamiltonian system arising from the singular lqr problem, in: Proceedings
  of the 21st IFAC World Congress, Berlin, Germany, 2020.

\bibitem{WilKitSil:86}
J.~Willems, A.~Kitap{\c{c}}i, L.~Silverman, Singular optimal control: a
  geometric approach, {SIAM} {J}ournal on {C}ontrol and {O}ptimization 24~(2)
  (1986) 323--337.

\bibitem{WilTre:98}
J.~Willems, H.~Trentelman, On quadratic differential forms, {SIAM} {J}ournal on
  {C}ontrol and {O}ptimization 36~(5) (1998) 1703--1749.

\end{thebibliography}

\end{document}